# Band Modulation for Silicene and Graphene Quantum Dots: A First-Principles Calculation

Bi-Ru Wu

*Department of Natural Science, Center for General Education, Chang Gung University, Taiwan – brwu@mail.cgu.edu.tw*

**Abstract** – The band modulation of the silicene and graphene quantum dots is investigated by a first-principles method. This study includes the ordinary silicene and graphene quantum dots and the embedded quantum dots in the hydrogenated silicene and graphene. The shapes and sizes of quantum dots are recognized as important factors for the electronic properties. We studied several types of quantum dots: triangular, parallelogram, rectangular, hexagonal dots. It demonstrates the energy gap of the quantum dot can be tuned by the dot size, the larger of the dot the smaller the energy gap. Moreover, the shapes affect the magnetism of the quantum dots. The triangular dot exhibits as magnetic semiconductor; the parallelogram dot shows antiferromagnetic characteristics; while the hexagonal dot is non-magnetic. Control the size and shape of a silicene or graphene quantum dot can manipulate its magnetism and electronic properties.

**Introduction**

Quantum dots (QDs) have drawn many research interests because of its abundant physical properties, flexibility in designing and wide application, such as optical devices and biosensing [1-6]. It is well known that QDs have highly tunable characteristics and attracts wide attention [3-6]. Their electronic properties changes as a function of both size and shape [4-6]. As our finding that the shape of a quantum dot (QD) not only changes the electronic structure but also the magnetism. In this paper, the graphene and silicone QDs I present contains the ordinary QDs and the embedded QDs in hydrogenated graphene and silicene. The hydrogenated graphene is called graphane and the hydrogenated silicene is named as silicane [7-9].

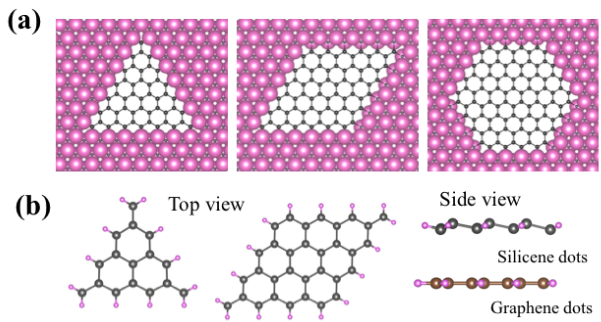

**Figure 1** –(a) The embedded QDs, and (b) top view and side view of the ordinary QDs. Pink balls denote the hydrogen atoms and black balls are corresponding to the silicon and carbon atoms in silicene and graphene dots, respectively.

The embedded QDs as shown in Fig.1a are the region of missing hydrogen atoms in silicane or graphane. The pink balls denote hydrogen atoms; they are enlarged to see the shape of QDs clearly. These regions can be a triangle, a parallelogram, or a hexagon. The triangular shaped dot is called embedded triangular QD, the name for the QD of other shapes follows the same rule. The embedded triangular, parallelogram and hexagonal dots are all in zigzag edges. The ordinary QDs with the same shaped are also calculated. The ordinary silicene and graphene QDs (Fig. 1b), their edges are passivated by hydrogen atoms and all in zigzag edges. The graphene dots are planar but the silicene dots are buckled in the dot plan.

*Method*

This calculation is based on the density functional theory [10] within the generalized gradient approximation, the projector-augmented wave method is used. A super cell is selected to simulate an individual QDs; 10 Å is used for the lateral distance between the edge of two dots and the vacuum distance in the perpendicular direction is 20 Å to avoid the interaction between the nearest super cell. The VASP code was used in the calculation. [11,12]

**Calculation details**
*Structure*

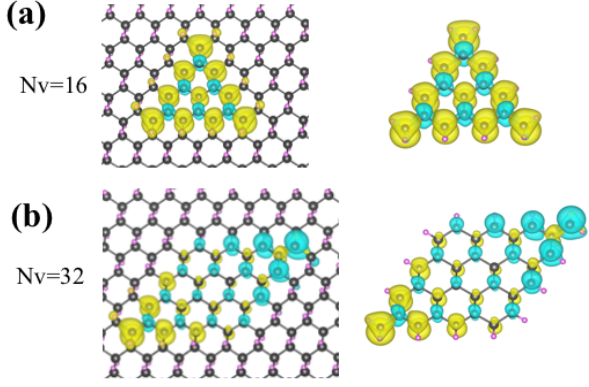

**Figure 2** –Spin charge density (SCD) of the embedded silicene QDs (Left) and ordinary silicene QDs (Right). (a) SCD of triangular QDs, and (b) SCD of parallelogram QDs. Yellow bobbles and blue bobbles are the spin up

## Results and Discussion
*Magnetism*
The spin charge density (SCD) reveals, all embedded triangular dots are magnetic (Fig.2a), the spin orientation of atoms at A and B sites are opposite. The atom at three tips have the largest magnetic moment, and the magnetic moment of edge atoms is lager than that of inner atom. The SCD distribution of an ordinary triangular dot is similar to the embedded one. However, the magnetic moment at three tips of an ordinary dot is slightly larger than that of the embedded one. Magnetic moment of the atom at tip of triangular graphene dot is slightly larger than that of the silicene dot. The total magnetic moment of silicene and graphene dots increases as the number of Si and C atoms, and is nearly root square of the number of atoms (Fig. 3). The increasing rate of total magnetic moment of embedded silicene QD is less than that of embedded graphene QD and ordinary silicene QD.

The parallelogram QDs show as antiferromagnetic semiconductor except for the small graphene QDs with eight atoms whether the ordinary or embedded QD. It is nonmagnetic semiconductor. Atoms at two tips of the parallelogram dot have the largest magnetic moment but opposite spin orientation. The Si (C) atoms have smaller moments depending on the spatial distances from the tips. The magnetic moment of tips in graphene QD is larger than that in silicene QD.

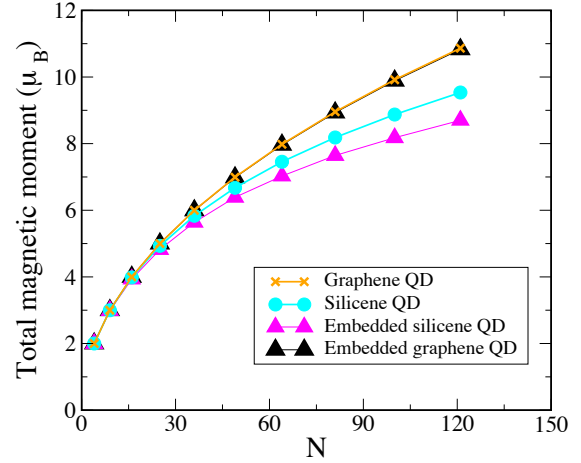

**Figure 3** –Total magnet moment of four types triangular QDs

*Electronic structure*
The flat bands in the wide band gap of silicane are contributed by the bare Si atoms inside the QD. The spin orientation of the flat valence bands and conduction band of triangular QDs are opposite (Fig. 4a). The band structure of the embedded parallelogram silicene QDs also reveals the antiferromagnetic characteristic (Fig. 4b). The bands with opposite spin orientation are degenerate. The flat bands are also in the wide energy gap of silicane. The energy gap decreases as the size of QD increases (Fig. 5). The energy gap of silicene dot is smaller than that of graphene dot. The gap of an ordinary QD is larger than the embedded one.

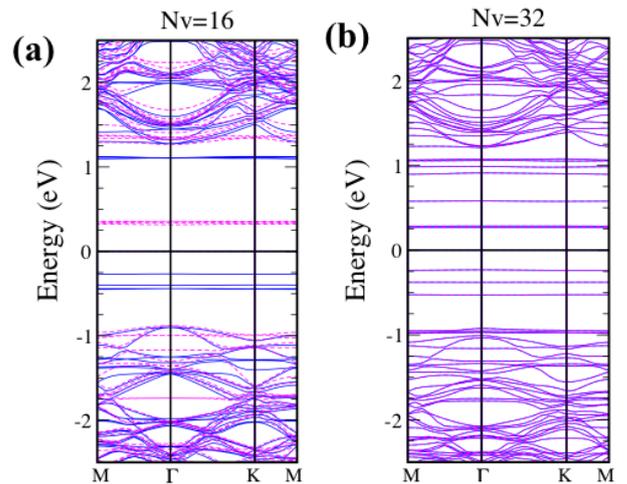

**Figure 4** – Band structure of embedded silicene QDs for (a) triangular and (b) parallelogram shapes. Blue solid lines are the bands of majority spin, while pink lines are the bands of minority spins.

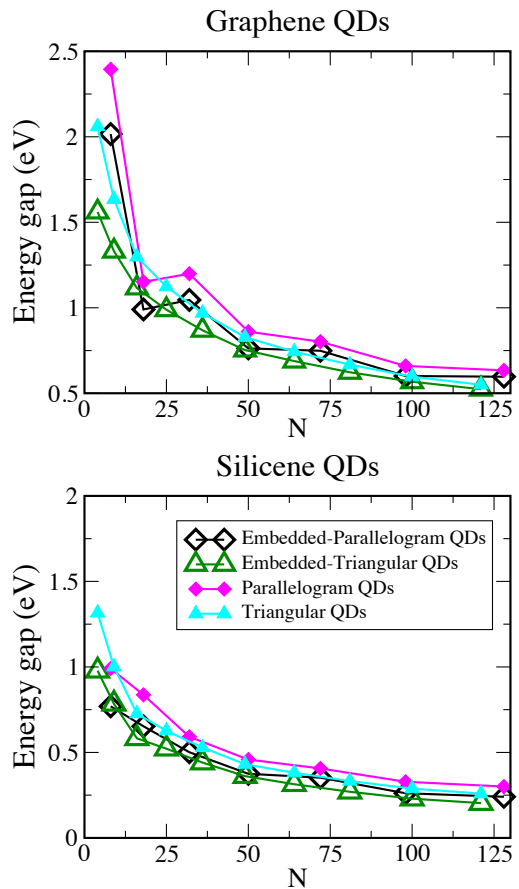

**Figure 5** – Energy gap for the (a) graphene QDs and (b) silicene QDs

**Conclusions**

The magnetic properties of both embedded and ordinary QDs determined by the shape of QD. All triangular QDs are magnetic semiconductors, while the parallelogram dots are antiferromagnetic. The energy gap decreases as the size of the QD is enlarged.

**Acknowledgements**

This work was supported by the Ministry of Science and Technology of the Republic of China under grant numbers MOST 105-2112-M-182-002-MY3. Supports from the National Centers for Theoretical Sciences and High-performance Computing of the ROC are also gratefully acknowledged.